\documentclass[12pt]{article} 

\renewcommand{\thefootnote}{\fnsymbol{footnote}}

\usepackage{comment}
\usepackage{latexsym}
\usepackage{amssymb,amsfonts}

\usepackage[utf8]{inputenc}
\usepackage[T1]{fontenc} 
\usepackage{lmodern}
\usepackage[english]{babel}           

\usepackage[pdftex]{graphicx}
\usepackage{epsfig}
\usepackage{graphicx}
\usepackage{comment}
\usepackage{latexsym}
\usepackage{hyperref}
\usepackage{amsmath}
\usepackage{mathrsfs}
\usepackage[usenames, dvipsnames]{color}
\usepackage{amsbsy}
\usepackage{amssymb}
\usepackage{amsthm}
\usepackage{amsfonts}
\usepackage{cite}
\usepackage{enumitem}
\usepackage{xcolor}
\usepackage{color}
\usepackage{mathtools}
\usepackage{caption} 
\usepackage{subcaption} 
\usepackage{esvect}
\usepackage{cancel}

\usepackage{diagbox}

\usepackage{tikz}

\usepackage{tcolorbox}

\newcommand{\be}{\begin{equation}}
\newcommand{\ee}{\end{equation}}
\newcommand{\bea}{\begin{eqnarray}}
\newcommand{\eea}{\end{eqnarray}}

\newcommand{\dd}{\text{d}}

\newcommand{\bm}{\boldmath} 

\newcommand{\Vone}{V_{\mbox{\scriptsize 1-loop}}}
\newcommand{\Z}{{\mathbb Z}}
\newcommand{\F}{{\cal F}_{\mbox{\scriptsize 1-loop}}}

\catcode`\@=11
\def\marginnote#1{}
\newcount\hour
\newcount\minute
\newtoks\amorpm
\hour=\time\divide\hour by60 \minute=\time{\multiply\hour by60
\global\advance\minute by-\hour}
\edef\standardtime{{\ifnum\hour<12 \global\amorpm={am}%
        \else\global\amorpm={pm}\advance\hour by-12 \fi
        \ifnum\hour=0 \hour=12 \fi
        \number\hour:\ifnum\minute<10 0\fi\number\minute\the\amorpm}}
\edef\militarytime{\number\hour:\ifnum\minute<10 0\fi\number\minute}
\def\draftlabel#1{{\@bsphack\if@filesw {\let\thepage\relax
   \xdef\@gtempa{\write\@auxout{\string
      \newlabel{#1}{{\@currentlabel}{\thepage}}}}}\@gtempa
   \if@nobreak \ifvmode\nobreak\fi\fi\fi\@esphack}
        \gdef\@eqnlabel{#1}}
\def\@eqnlabel{}
\def\@vacuum{}
\def\draftmarginnote#1{\marginpar{\raggedright\scriptsize\tt#1}}
\def\draft{\oddsidemargin -.2truein
        \def\@oddfoot{\sl preliminary draft \hfil
        \rm\thepage\hfil\sl\today\quad\militarytime}
        \let\@evenfoot\@oddfoot \overfullrule 3pt
        \let\label=\draftlabel
        \let\marginnote=\draftmarginnote
   \def\@eqnnum{(\theequation)\rlap{\kern\marginparsep\tt\@eqnlabel}%
\global\let\@eqnlabel\@vacuum}  }
\def\thebibliography#1{
\vskip 0.5cm \centerline{\bf \Large References}
\list{
[\arabic{enumi}]}{\settowidth\labelwidth{[#1]}
\leftmargin\labelwidth
\advance\leftmargin\labelsep
\usecounter{enumi}}
\def\newblock{\hskip .11em plus .33em minus .07em}
\sloppy\clubpenalty4000\widowpenalty4000
\sfcode`\.=1000\relax}

\renewcommand{\theequation}{\arabic{section}.\arabic{equation}}
\renewcommand{\section}{\setcounter{equation}{0}\@startsection
{section}{1}{0mm}{-\baselineskip}{0.5\baselineskip} {\normalfont\Large\bfseries}}
\renewcommand{\subsection}{\@startsection
{subsection}{2}{0mm}{-\baselineskip}{0.5\baselineskip} {\normalfont\large\bfseries}}
\renewcommand{\subsubsection}{\@startsection
{subsubsection}{3}{0mm}{-\baselineskip}{0.5\baselineskip}
{\normalfont\normalsize\slshape}}

\begin{document}


\begin{titlepage}
\begin{flushright}
CPHT-PC104.122019, December   2019
\vspace{1cm}
\end{flushright}
\begin{centering}
{\bm\bf \Large Spontaneous Freeze Out of Dark Matter\footnote{Based on a talk given at the ``Conference on Recent Developments in Strings and Gravity'', 10--16 September 2019, Corfu, Greece.}}

\vspace{7mm}

 {\bf Thibaut Coudarchet\textsuperscript{1}, Lucien Heurtier\textsuperscript{2} and Herv\'e Partouche\textsuperscript{1}}

 \vspace{4mm}

$^1$CPHT, CNRS, Ecole polytechnique, IP Paris, F-91128 Palaiseau, France\\
{\em  thibaut.coudarchet@polytechnique.edu,\\
herve.partouche@polytechnique.edu}

$^2${Department of Physics, University of Arizona, Tucson, AZ   85721\\
{\em heurtier@email.arizona.edu}}

\end{centering}
\vspace{0.7cm}
$~$\\
\centerline{\bf\Large Abstract}\\

\begin{quote}
We present a new paradigm for the production of dark-matter particles called the spontaneous freeze out, in which the decoupling from the thermal bath is enforced by the sudden increase of the dark-matter mass, due to the spontaneous breaking of some global symmetry rather than by the slow decrease of the temperature. We study the details of the spontaneous freeze out mechanism using a simple toy model and analyze the phenomenology of our scenario. We show that in order to obtain the correct relic abundance, the annihilation cross section of dark-matter particles into Standard-Model states has to be typically much larger than in the case of a constant-mass thermal dark-matter candidate. We present a string theory model in which such a scenario takes place naturally. 
\end{quote}

\end{titlepage}
\newpage
\setcounter{footnote}{0}
\renewcommand{\thefootnote}{\arabic{footnote}}
 \setlength{\baselineskip}{.7cm} \setlength{\parskip}{.2cm}

\setcounter{section}{0}


\section{Introduction}
The Weakly Interacting Massive Particle (WIMP) paradigm, standing among the most popular production mechanism for dark-matter (DM) particles in the early universe relies on the assumption that the mass of the DM particles is constant throughout the evolution of the universe. However, it is well known that the presence of a thermal equilibrium at early stages of cosmology can affect the stabilization of scalar fields and lead to a temperature-dependent behaviour of fermion masses as the temperature goes down. If dark matter is a fermionic particle acquiring its mass through the spontaneous breaking of some ultra-violet global symmetry, we show that the dynamics of the phase transition leading to such breaking can affect significantly the freeze out mechanism. We first detail the main features of the spontaneous mechanism on a simple toy model, and then present a well motivated string theory model in which such mechanism naturally takes place. Further details beyond the results	presented here can be found in Refs.~\cite{SFO, SFOstring}

\section{The spontaneous freeze out mechanism}

The simplest dark sector that may yield a spontaneous freeze out is composed of  a Dirac or Majorana dark matter fermion $\psi$, and  a real scalar field $\phi$. Due to a Yukawa coupling, the mass of $\psi$ is determined by the vacuum expectation value (vev) of $\phi$.  The renormalizable classical Lagrangian density $\mathcal L_{\rm dark}$ in the dark sector is assumed to admit a $\Z_2$ symmetry $\phi\to -\phi$. 
To take place, the mechanism requires the tree-level mass term of the scalar field to be negative, for the classical scalar vev not to vanish. Hence, we define 
\be
\mathcal L_{\rm dark}=i\bar \psi\cancel{\partial}\psi+\frac{1}{2}\partial_\mu \phi\partial^\mu \phi-y \phi \bar\psi\psi -V_{\rm tree}(\phi)\,,
\label{lag}
\ee 
where the scalar potential takes the familiar form\footnote{In our conventions, $y$, $\mu$ and $\lambda$ are all positive. }
\be V_{\rm tree}(\phi)=-\frac{\mu^2}{2} \phi^2+\frac{\lambda}{4!}\phi^4\,.
\ee
After reheating, the Standard Model (SM) particles are in thermal equilibrium at a temperature $T$, and we assume the existence of an epoch during which interactions between the dark and  the visible sectors are strong enough to maintain $\psi$ and $\phi$ thermalized with the SM bath. These interactions will be specified in the sequel. 

At the quantum level, it turns out that the effective potential at finite temperature of $\phi$ restores at high temperature the $\Z_2$  symmetry of the vacuum~\cite{Quiros:1999jp}. Hence, in this regime, thermal loop corrections to the potential dominate over the tree-level contribution, implying  perturbation theory to breakdown. Diagrams with an arbitrary number of loops, but still of the same order of magnitude, must be taken into account. Their dominant effects arise from the so-called ``ring diagrams'' (or ``daisy diagrams''), from which the high temperature limit can be extracted~\cite{Dolan:1973qd,Carrington:1991hz}. In total, a consistent expression of the thermal effective potential is found by taking into account three types of quantum corrections,
\be
V^{\rm th}_{\rm eff}(T,\phi)=V_{\rm tree}(\phi)+\Vone(\phi)+\F(T,\phi)+V_{\rm ring}^{\rm th}(T,\phi)\,,
\label{Veff1}
\ee
which can be described as follows:
\begin{itemize}

\item $\Vone(\phi)$ is the zero-temperature Coleman--Weinberg effective potential at 1-loop. By adjusting appropriate counter-terms  in the $\overline{\rm MS}$ scheme, it can be written as 
\be
\Vone(\phi)= \frac{m_0(\phi)^4}{64\pi^2}\left[\log\!\left(\frac{m_0(\phi)^2}{Q^2}\right)-\frac{3}{2}\right]-n_F\frac{m_\psi(\phi)^4}{64\pi^2}\left[\log\!\left(\frac{m_\psi(\phi)^2}{Q^2}\right)\!-\frac{3}{2}\right],
\ee
where $n_F=4$ ($n_F=2$) for a Dirac (Majorana) fermion $\psi$. In this formula, the bosonic and fermionic masses squared are given by
\be
m_0(\phi)^2=-\mu^2+{\lambda\over 2}\phi^2\, ,\qquad m_\psi(\phi)=y\phi\, ,
\label{m00}
\ee 
while $Q$ is the renormalization scale. 

\item $\F(T,\phi)$ is the Helmholtz free energy at 1-loop that is associated with the gas of particles $\psi$ and $\phi$. It can be expressed as 
\be
\F(T,\phi)=\frac{T^4}{2\pi^2}\!\left[J_B\!\left(\frac{m_0(\phi)^2}{T^2}\right)-n_F\, J_F\!\left(\frac{m_\psi(\phi)^2}{T^2}\right)\right],
\ee
where $J_B$ and  $J_F$ are functions defined as
\be
\label{JBF}
J_{B,F}\!\left(\frac{m^2}{T^2}\right) =\int_0^{+\infty}\dd u\,  u^2 \log \!\left(1\mp e^{-\sqrt{u^2+m^2/T^2}}\right).
\ee

\item Finally, $V_{\rm ring}^{\rm th}(T,\phi)$ is the higher-loop contribution arising from the ring diagrams. It takes the form~\cite{Delaunay:2007wb}
\be
V_{\rm ring}^{\rm th}(T,\phi)={T\over 12\pi}\Big[ \big(m_0(\phi)^2\big)^{3\over 2}-\big(m_0(\phi)^2+\Pi_\phi(T)\big)^{3\over 2}\Big],
\ee
in terms of the so-called Debye mass squared
\be
\Pi_\phi(T)= {T^2\over 24}(\lambda+n_Fy^2)\, , 
\ee
which is the dominant monomial at large $T$ of $\displaystyle\frac{\partial^2 \F}{\partial\phi^2}$. 

\end{itemize}
It turns out to be relevant to consider the high temperature expansion of  the function $J_B$,  which is given by 
\be
J_B\!\left(\frac{m^2}{T^2}\right)\!=-\frac{\pi^4}{45}+\frac{\pi^2}{12}\frac{m^2}{T^2}-{\pi\over 6}\Big({m^2\over T^2}\Big)^{3\over 2}-\frac{1}{32}\frac{m^4}{T^4}\log \!\left(\frac{m^2}{16\alpha T^2}\right)+\mathcal{O}\!\left(\frac{m^6}{T^6}\right),
\ee 
where $\alpha=\pi^2\exp(3/2-2\gamma_{\rm E})$ and $\gamma_{\rm E}$ is the Euler--Mascheroni constant. In this form, one can see that the $T^4\log(m_0^2)$ terms appearing in the 1-loop contributions $\Vone$ and $\F$ cancel exactly. Moreover, the terms $T(m_0^2)^{3\over 2}$ present in $\F$ and $V_{\rm ring}^{\rm th}$ also cancel one another. It is in that sense that the thermal effective potential given in Eq.~(\ref{Veff1}) is consistent at high temperature, since all terms that may source imaginary parts to $V_{\rm ring}^{\rm th}(T,\phi)$ when $m_0(\phi)^2<0$ are actually not present. Expanding $J_F$ in a similar way, 
\be
J_F\!\left(\frac{m^2}{T^2}\right) \!=\frac{7\pi^4}{360}-\frac{\pi^2}{24}\frac{m^2}{T^2}-\frac{1}{32}\frac{m^4}{T^4}\log \!\left(\frac{m^2}{\alpha T^2}\right)+\mathcal{O}\!\left(\frac{m^6}{T^6}\right),
\ee
one observes that the $T^4\log(m_\psi^2)$ terms also disappear from the 1-loop contribution $\Vone+\F$.

When neglecting all terms $T^4\times \mathcal{O}(m^6/T^6)$, the thermal effective potential can be formatted in a very suggestive way. For this purpose, we make a suitable choice of renormalization scale $Q$ and parameterize all dependencies on temperature with a new variable $x$, 
\be
Q=\pi e^{-\gamma_{\rm E}} T_c\, ,\quad\;\; x\equiv {T_c\over T}\, ,
\ee
where we have defined the following  critical temperature,
\be
T_c={2\sqrt{6}\, \mu\over \sqrt{\lambda+n_F y^2}}\, \sqrt{1-{\sqrt{6}\over 8\pi}\,\xi+{\log 2\over 8\pi^2}\,\lambda\over 1-{\sqrt{6}\over 4\pi}\,\xi}\,,\;\;\;\;\mbox{with}\;\;\;\;
\xi\equiv {\lambda\over \sqrt{\lambda+n_Fy^2}}\, .
\ee
In these conventions and notations, we obtain
\be
V^{\rm th}_{\rm eff}(x,\phi)=V_0(x)-\frac{\mu_{\rm eff}(x)^2}{2}\phi^2+\frac{\lambda_{\rm eff}(x)}{4!}\phi^4\,,
\ee
where $V_0$ is an irrelevant contribution independent on the scalar $\phi$, while  $\mu_{\rm eff}^2$ and $\lambda_{\rm eff}$ are effective mass terms and self-couplings:
\be
\begin{aligned}
\mu_{{\rm eff}}(x)^{2}&=\mu^2\bigg[\!\bigg(1-{\sqrt{6}\over 8\pi}\,\xi+{\log 2\over 8\pi^2}\,\lambda\bigg)\Big(1-{1\over x^2}\Big)-{\lambda\over 16\pi^2}\log x\bigg],\\
\lambda_{{\rm eff}}(x)&=\lambda\bigg(1-{3\sqrt{6}\over 8\pi}\,\xi+{3\log 2\over 8\pi^2}\,\lambda\bigg)+{3\over 16\pi^2}\big(4n_F y^4-\lambda^2\big)\log x\, .
\end{aligned}
\label{mleff}
\ee
The key point is that $\mu_{{\rm eff}}^{2}$ changes sign at the critical temperature ($x=1$), where a second order  phase transition takes place. The value of $\mu_{{\rm eff}}^{2}$ is negative at higher temperature, implying $\phi$ to be stabilized at zero, while it is positive below $T_c$, which triggers the Brout-Englert-Higgs mechanism, \textit{i.e.} the condensation of the scalar field. Hence, as the universe expands and the temperature drops, the vev of $\phi$ varies, and so do the thermally corrected masses in the dark sector,
\bea
x\le 1 :&&\!\!\langle\phi\rangle = 0\,,\qquad \qquad \quad\;\;\;\,\quad m_{\phi}(x) = |\mu_{{\rm eff}}(x)|\, , \quad \;\;\,m_{\psi}(x) = 0\, ,\label{mevol}\\
x\ge 1 :&&\!\!\langle\phi\rangle = \mu_{{\rm eff}}(x)\sqrt{\frac{6}{\lambda_{\rm eff}(x)}}\,,\;\;\; m_{\phi}(x) = \sqrt{2} \,\mu_{\rm eff}(x)\, , \;\;\; m_{\psi}(x) \equiv y\langle \phi\rangle=  y\sqrt{\frac{3}{\lambda_{\rm eff}(x)}}\, m_{\phi}(x)\, .\nonumber 
\eea

The relations derived so far are actually valid until one of the species $\psi$ or $\phi$ decouples from the thermal bath. If several scenarios may be considered, we will concentrate in the sequel on the case where the dark fermion is maintained in thermal equilibrium with the SM because of contact interactions between $\psi$ and the SM fields. Therefore, the dark-matter fermion $\psi$ freezes out at some $x=x_{\rm FO}\equiv T_c/T_{\rm FO}$, when the condition 
\be
H<n_\psi\langle \sigma_{{\rm SM}\leftrightarrow\psi\bar\psi} \,v\rangle
\ee
ceases to be satisfied. In our notations, $\langle \sigma_{{\rm SM}\leftrightarrow\psi\bar\psi} \,v\rangle$ is the thermally averaged cross section of annihilation of $\psi$ into SM particles, $n_\psi$ is the number density of $\psi$, and $H$ stands for the Hubble parameter. We are interested in the regime where the freeze out of the dark-matter fermion arises because of its mass increase, so that $x_{\rm FO}>1$. 
We will also assume that the scalar $\phi$ remains in thermal equilibrium with the SM bath even when the particles $\psi$ have frozen out. Before the phase transition occurring at $x=1$, thermalization of $\phi$ can be accounted for by the inverse decay $\psi+\bar \psi\to \phi$. After the phase transition, interactions between $\phi$ and SM fields can maintain $\phi$ in equilibrium, as will be seen in the next section in specific examples.

In order to characterize regions in the parameter space with distinct features, it is useful to introduce the ratio
\be
\kappa={m_\psi(x_{\rm FO})\over T_{\rm FO}}\, .
\ee
In fact, even if $\kappa$ depends on the details of the annihilation cross section, it is in practice ${\cal O} (20 \mbox{--} 30)$ for DM masses in the GeV--TeV range. From this definition, it is straightforward to show that 
\be
\begin{aligned}
\lambda\gg n_Fy^2\quad& \Longrightarrow\quad x_{\rm FO}\simeq  {\cal O}\Big({2\kappa\over y}\Big)\gg \kappa\, ,  \\
\lambda\ll n_Fy^2 \quad &\Longrightarrow \quad  x_{\rm FO}\simeq \left[1+\kappa^2\Big({4 \lambda\over n_F y^4}+{3\over \pi^2}\log x_{\rm FO}\Big)\right]^{1/2}\, ,
\end{aligned}
\ee
from which we conclude  that $x_{\rm FO}$ is relatively smaller than $\kappa$ when $\lambda < y^4$. To proceed, it is enough to extract from the results displayed in Eq.~(\ref{mevol}) the evolutions at leading order in small couplings of the masses in the dark sector, 
\be
1\leqslant x\leqslant x_{\rm FO} :\quad m_{\phi}(x) \simeq \mu\sqrt{\frac{6}{\lambda}} \sqrt{1-{1\over x^2}}\,, \quad 
m_{\psi}(x) \simeq  y \mu\sqrt{{6\over \lambda+{3n_F\over 4\pi^2}y^4\log x}}\,\sqrt{1-{1\over x^2}} \,.
\ee
From these formulas, we distinguish two different regimes:
\begin{itemize}
\item When $x_{\rm FO}\gg \kappa$, we have just seen that the term $y^4\ll \lambda$ in the expression of $m_\psi$ can be omitted. Moreover, the temperature when DM freezes out is so small, compared to $T_c$, that the masses of $\phi$ and $\psi$ have already reached their  asymptotic values.  Therefore, this case corresponds to the usual freeze out scenario, where the DM particle $\psi$ has a constant mass. 

\item On the contrary, when $x_{\rm FO}\lesssim  \kappa$ (\textit{i.e.} $\lambda < y^4$), the masses in the dark sector are still evolving when the decoupling takes place. In fact, the vev of the scalar $\phi$ at $x_{\rm FO}$ is still far from its final value at present time. As a result, a significant increase of the mass of $\psi$ occurs between the time at which it decouples from the thermal bath and today. When the parameters of the model lead to this  regime, we will refer to the decoupling of DM as a Spontaneous Freeze-Out.  

\end{itemize}

After DM freezes out, the fact that the masses of the particles $\psi$ depend on the vev of $\phi$ implies that the relic density contributes to the scalar potential.   In particular, when the temperature of the universe is much lower than the mass of $\phi$ at zero temperature, the scalar potential takes the form
\be
V_{\rm tot}(\phi)=V_{\rm tree}(\phi)+\Vone(\phi)+V_{\rm relic}(\phi)\, , 
\ee
where the last term stands for the contribution of the DM particles. By writing the equation of conservation of the stress-energy tensor for the system comprising the scalar $\phi$, the SM radiation and the relic density, one can show that~\cite{po1,po2,po3,po4,SFO} 
\be
x\gg x_{\rm FO} : \quad \dot \rho_{\rm relic}+3H(\rho_{\rm relic}+P_{\rm relic})=\dot \phi\,  {\dd V_{\rm relic}\over \dd\phi}\, ,
\label{conpart}
\ee
where $\rho_{\rm relic}, P_{\rm relic}$ are the energy density and pressure of the perfect fluid associated with the relic density. In our case of interest, the DM particles are non-relativistic. Neglecting for simplicity their velocities, the fluid is pressureless, its energy arises from the invariant mass, and its contribution to the potential of $\phi$ can be found from Eq.~(\ref{conpart}) and the fact that $n_\psi\propto 1/a^3$, where $a$ is the scale factor: 
\be\label{eq:dust}
\rho_{\rm relic}=n_\psi y |\phi|\,, \quad P_{\rm relic}=0\, ,\quad {\dd V_{\rm relic}\over \dd\phi}=\mathrm{sign}(\phi)y n_\psi \,.
\ee
However, due to the expansion of the universe (or the smallness of $n_\psi$ to account for the observed relic density), the correction to the potential arising from the dust can be neglected at present time. Doing so, a careful analysis of $V_{\rm tot}(\phi)$ leads to the conclusion that for a vacuum at some $\langle \phi\rangle>0$ to exist, $\lambda/y^4$ should be greater than a lower bound,
\begin{equation}\label{eq:bound}
\frac{\lambda}{y^4}>\frac{3 n_F}{8\pi^2}\left(\log\frac{2}{3}+2\gamma_{\rm E}\right)\simeq 0.03\, n_F\,.
\end{equation}
As a result, the SFO takes place when $1<x_{\rm FO}\lesssim \kappa$, while having $1\lesssim x_{\rm FO}< \kappa$ is excluded. 


\section{Phenomenological consequences}

\begin{figure}
\begin{center}
\includegraphics[width=0.8\linewidth]{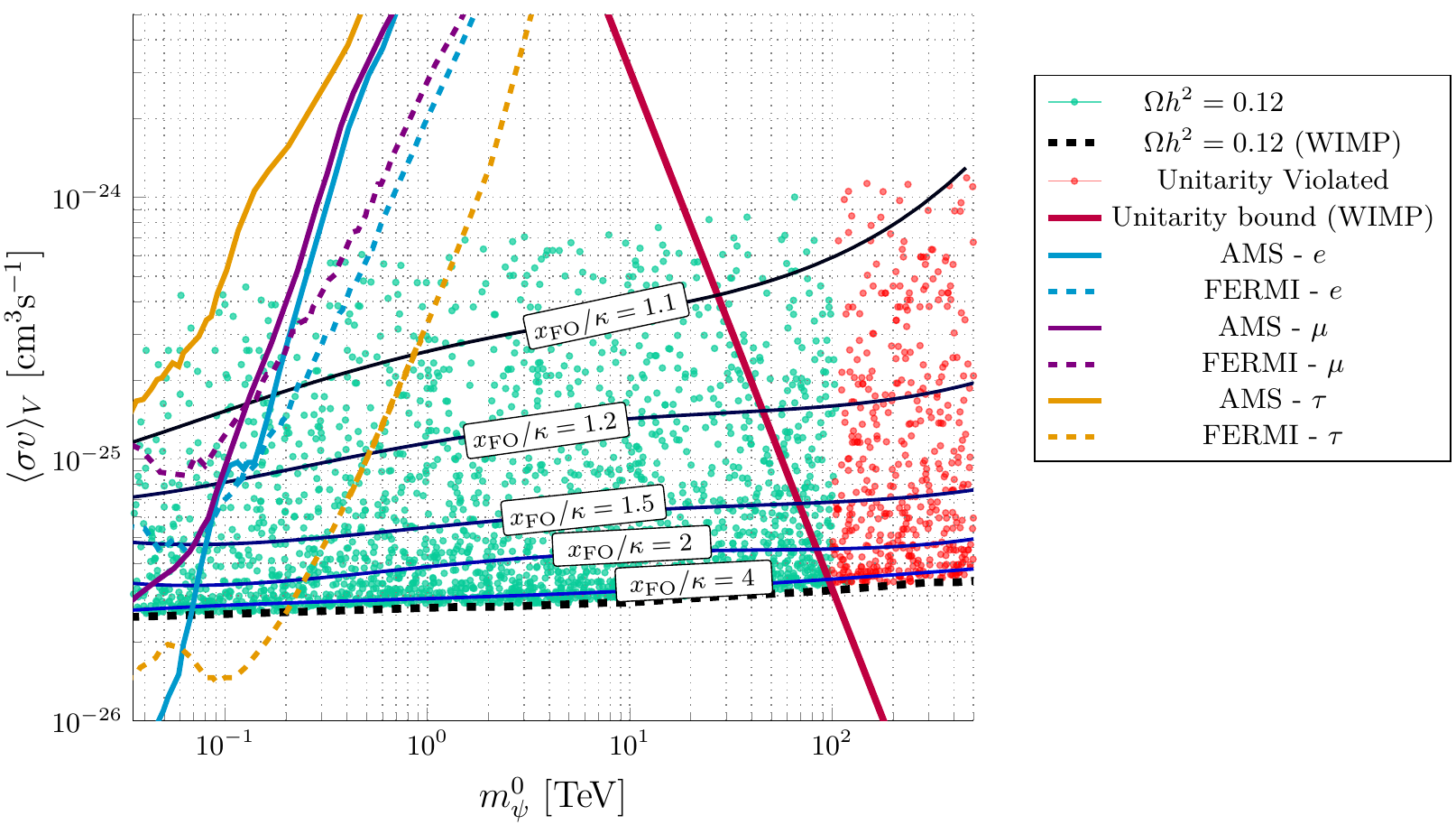}
\end{center}
\caption{\label{fig1}  \footnotesize Numerical results for the cross section of annihilation of dark-matter particles interacting with SM fermions \textit{via} the operator $\mathcal{O}_V$. The plain red line indicates the standard WIMP unitarity bound, whereas the red dots stand for the points which violate unitarity in our scenario. Indirect-detection constraints on the annihilation cross section of a dark-matter candidate interacting with one single species of lepton are indicated.}
\end{figure}

\begin{figure}
\begin{center}
\includegraphics[width=0.8\linewidth]{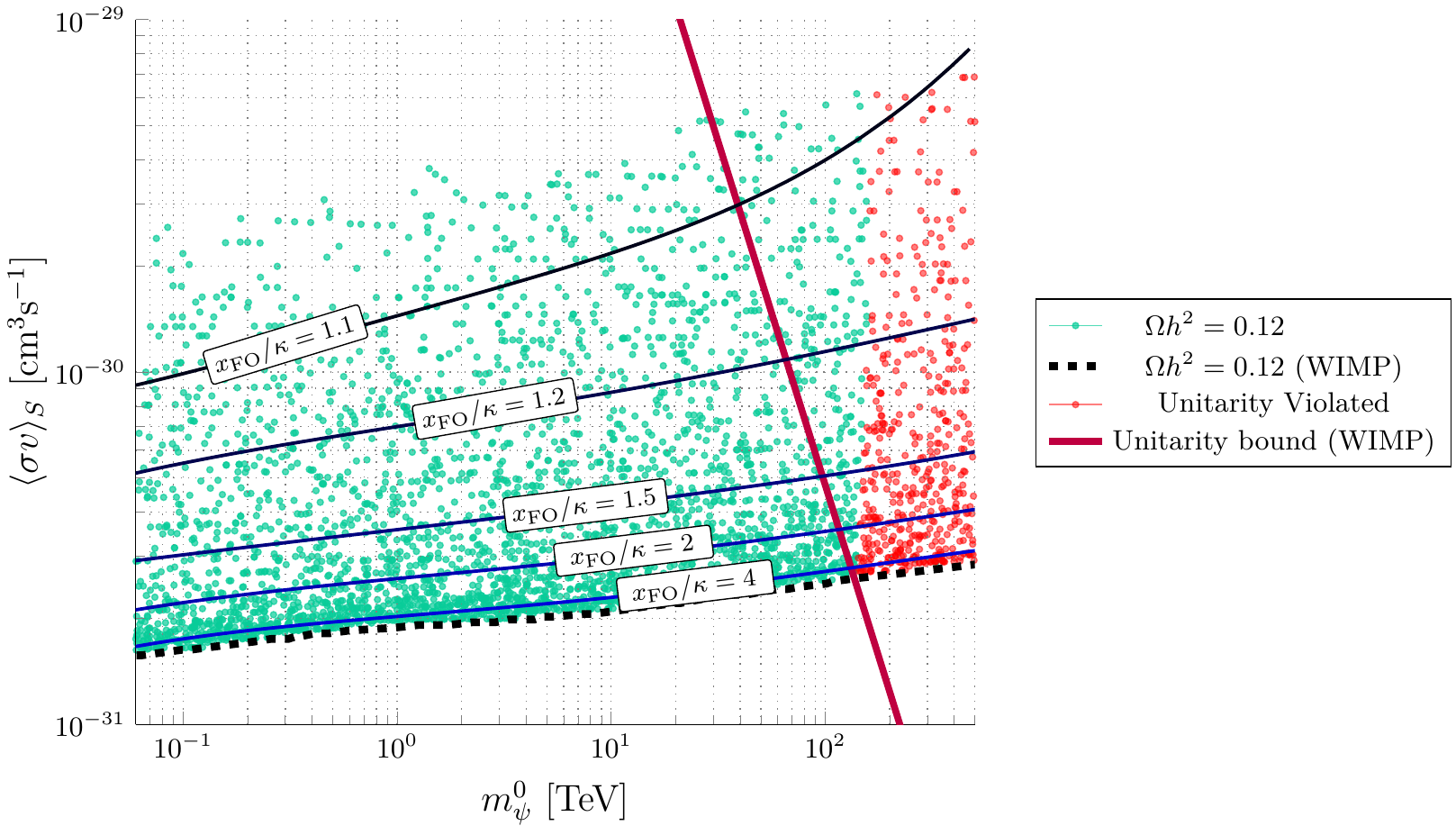}
\end{center}
\caption{\label{fig2}  \footnotesize Numerical results for the cross section of annihilation of dark-matter particles interacting with SM fermions \textit{via} the operator $\mathcal{O}_S$. The plain red line indicates the standard WIMP unitarity bound, whereas the red dots stand for the points which violate unitarity in our scenario.}
\end{figure}

In order to study the phenomenological features of the SFO regime, we considered two benchmark operators encoding the interaction of DM particles with SM fermions, focusing on spin-independent interactions
\begin{equation}\label{eq:operators}
\mathcal O_V = \bar \psi {\gamma_\mu}\psi \bar f {\gamma^\mu}f\quad\text{and}\quad \mathcal O_S = \bar \psi\psi \bar f f\,,
\end{equation}
where $f$ can be any fermion of the SM or any fermion in thermal equilibrium with the SM at the time where DM particles freeze out. Using the evolution of the DM mass given in Eq.~(\ref{mevol}) the thermally-averaged annihilation cross section reads for these two operators
\begin{equation}
\begin{aligned}\label{eq:thermalXsection}
\langle \sigma v\rangle_V &\simeq  \frac{G_V^2}{2 \pi }\Big(1+\frac{x^{-1}T_c}{m_\psi(x)}\Big)\,m_\psi^2(x)\,,\\
\langle \sigma v\rangle_S&\simeq  \frac{3G_S^2 }{8 \pi }\, x^{-1} T_c m_\psi(x)\,. 
\end{aligned}
\end{equation}
We numerically solved the Boltzmann equation  for the yield $Y_\psi=n_\psi/s$, where $s$ is the entropy density of the visible sector, 
in order to derive the relic abundance of DM given such an evolution of the annihilation cross section. Denoting $Y_{\psi,\rm{eq}}$ the DM yield when DM particles follow a Boltzmann equilibrium, the equation can be written as 
\begin{equation}
\frac{\dd Y_\psi}{\dd x}=\frac{\langle \sigma v\rangle s}{x H}(Y_{\psi,\rm{eq}}^{2}-Y_\psi^2)\,.
\end{equation}
Scanning over the parameter space, we obtain for these two operators the results presented in Fig.~\ref{fig1} and \ref{fig2} in which we show the annihilation cross section of DM particle into SM fermions at present time that is found when the correct relic abundance is obtained. From these figures it is clear that points in the parameter space which correspond to low values of $\kappa/x_{\rm FO}$, and therefore to the SFO case, require an annihilation cross section of DM particles into SM fermions which can be larger by more than one order of magnitude than the cross section of annihilation required in the constant-mass WIMP case.
This consequence of the SFO regime can be understood using the following arguments:
\begin{itemize}
\item For a fixed dark-matter mass $m_\psi^0$ at present time and demanding the correct relic abundance of DM particles today, the quantity $Y_\psi^{\rm FO}$ is accordingly fixed.
\item The relative velocity of dark-matter particles at freeze out  \mbox{$\langle v^2\rangle\sim T_{\rm FO}/m_\psi=\kappa^{-1}$} is essentially model independent. Because in our scenario $m_\psi(x_{\rm FO})<m_\psi^0$, we expect the freeze-out temperature in the SFO case to be lower than in the constant-mass standard WIMP scenario.
\item At freeze out, the condition $n_\psi^{\rm FO}\langle\sigma v\rangle_{\rm FO} =H_{\rm FO}$ can be expressed as 
\begin{equation}
Y_\psi^{\rm FO}\langle\sigma v\rangle_{\rm FO}\propto T_{\rm FO}^{-1}\,.
\end{equation} 
Therefore the dark-matter annihilation cross section at freeze out $\sigma_{\rm FO}$ in the SFO case is larger than in the constant-mass paradigm.
\item In our benchmark models, the cross sections of Eq.~\eqref{eq:thermalXsection} evolve after freeze out as
\be
\begin{aligned}
\quad \langle \sigma v\rangle_V &\simeq  \frac{G_V^2}{2 \pi }\Big(1+\frac{v^2}{6}\Big)\,m_\psi^2(x)\,,\\
\quad \langle \sigma v\rangle_S&\simeq  \frac{3G_S^2 }{48 \pi }\, v^2 m_\psi(x)\,.
\end{aligned}
\ee
Because in the SFO case the dark-matter particle mass increases with time,
the ratio between the annihilation cross section in our scenario as compared to the usual WIMP case increases from the time of freeze out to present time, where the DM velocity $v\simeq 200~\mathrm{km.s^{-1}}$ is model independent.
\end{itemize}
This significant increase of the DM annihilation cross section typically renders DM particles more likely to be discovered experimentally than in the case of a constant-mass WIMP candidate. For this reason, a coupling of a dark fermion to coloured SM quarks for the two operators we considered are automatically ruled out in the SFO case. The SM fermion $f$ has therefore to be a standard model lepton or a new particle in equilibrium with the SM at the time of DM freeze out. In the case of the vectorial operator $\mathcal{O}_V$, leading to an $s$-wave annihilation cross section which could lead to visible signals in the galaxy, we compare our results to the experimental limits set on DM annihilation by AMS \cite{Aguilar:2014mma,Accardo:2014lma} and Fermi \cite{Ackermann:2015zua, Fermi-LAT:2016uux}. As one can see on Fig.~\ref{fig1}, a leptophillic DM candidate with a mass at the TeV scale could be detected in a near future by those experiments.

It is interesting to note that the evolution of the dark-matter mass and annihilation cross section between the time of freeze out and current time leads to a modification of the unitarity bound in the SFO case as compared to the standard freeze-out case. Indeed, particles of $\mathcal{O}(100)$ TeV and which constitute 100\% of the relic abundance might be able to interact with SM fermions with a larger cross section than it is allowed in the case of the constant-mass WIMP scenario.

\begin{figure}
\includegraphics[width=0.485\linewidth]{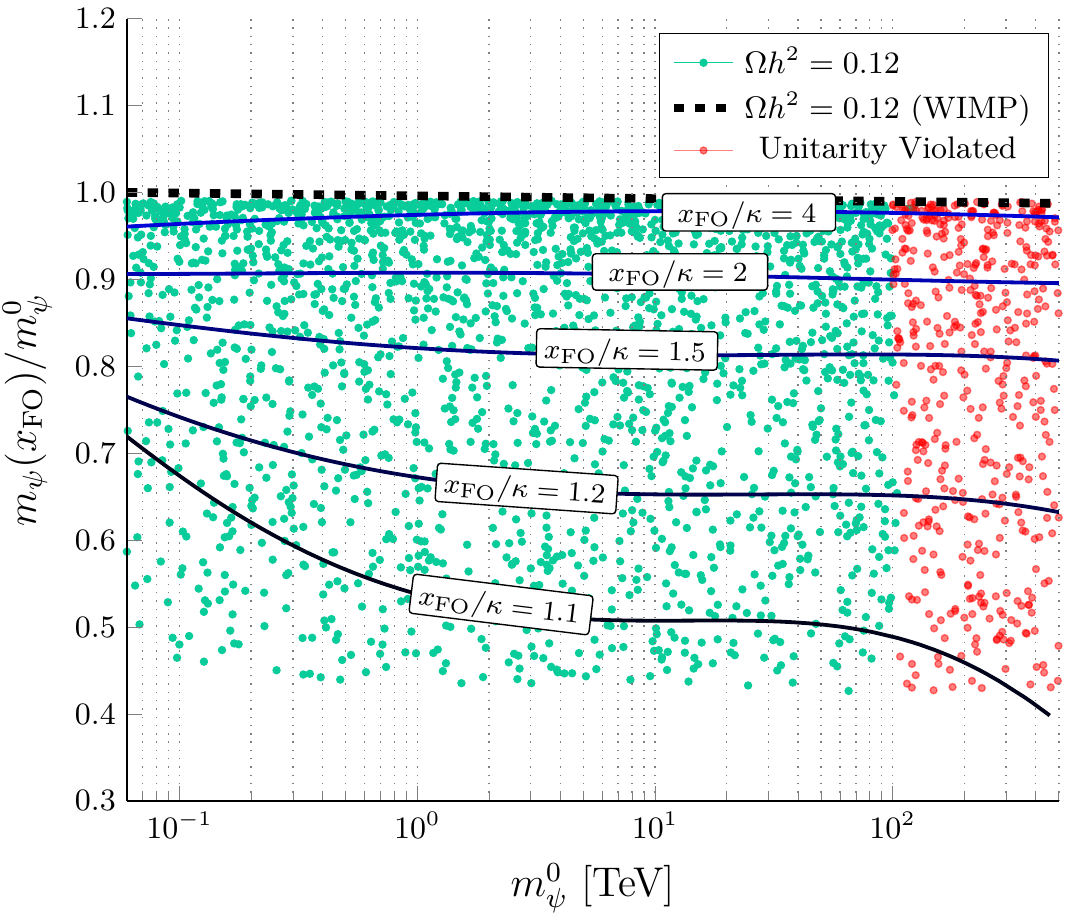}\hspace{0.02\linewidth}\includegraphics[width=0.485\linewidth]{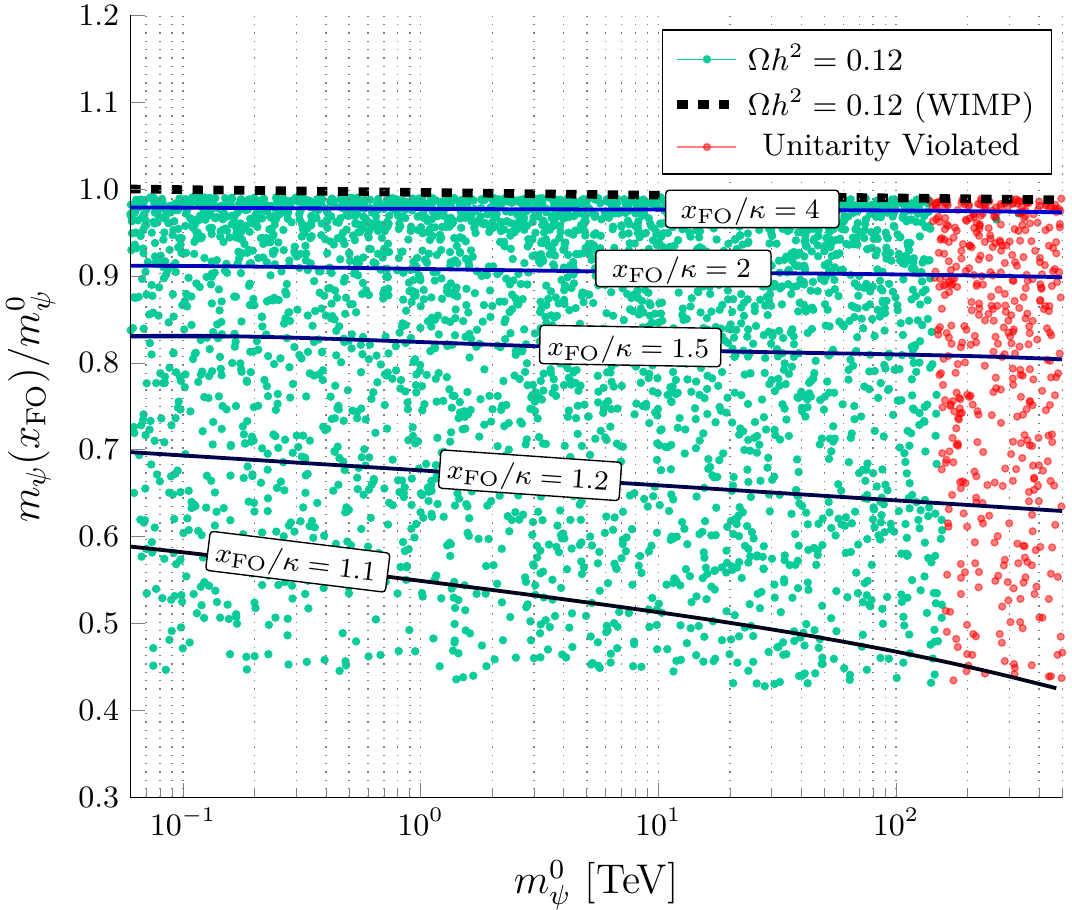}
\caption{\label{fig3}\footnotesize Ratio between the values of the dark-matter particle mass at the time of freeze-out and at present time for data points leading to the correct relic abundance of dark matter. The left panel corresponds to an $s$-wave annihilation cross section (operator $\mathcal O_V$), whereas the right panel stands for the $p$-wave annihilation cross section (operator $\mathcal O_S$). Red circles are associated with data points for which the annihilation cross sections violate unitarity at the time of freeze out.}
\end{figure}

We also show in Fig.~\ref{fig3} the ratio between the dark-matter mass at the time of freeze out and at present time. One can see that this ratio can be as small as 0.5. Below that value, as seen in Eq.~(\ref{eq:bound}), the Coleman Weinberg potential at zero temperature destabilizes the vacuum and the SFO mechanism cannot be realized anymore.

It is important to note that the presence of a decay term for the scalar is crucial for our scenario to be consistent. Indeed, after the phase transition takes place, in the case where $\lambda\ll n_F y^2$, whether we are in the spontaneous freeze out situation  (where this condition is always satisfied) or not, the ratio of the masses in the dark sector can be written as
\be
\label{eq:ratio}
1\leqslant x\leqslant x_{\rm FO} :\quad \frac{m_\psi(x)^2}{m_\phi(x)^2}\simeq\frac{3y^2}{\lambda+\frac{3n_F}{4\pi^2}y^4\log x}\gg 1\,.
\ee
As a result, in this region of the parameter space, the (inverse) decay process $\phi\leftrightarrow \psi+\bar \psi$ is kinematically forbidden. In the absence of a decay channel into SM fermions for the scalar field, it is possible that the oscillations of the scalar field after the phase transition come to dominate the energy density of the universe, since they would contribute significantly to the matter relic abundance, and could overclose the universe. In our scenario, we assumed that such a decay channel exist, either because a loop of DM particles can lead to a decay into SM fermions, or because one can introduce a coupling between the dark scalar and the SM Higgs boson.

\section{A string theory approach}

Actually, the idea that thermalized DM acquires its mass \textit{via} condensation of a dark scalar and suddenly freezes out from the thermal bath was first introduced in the context of string theory~\cite{SFOstring}. In this section, our goal is to review how the thermal phase transition responsible for the DM mass generation can naturally occur during the universe evolution, in the context of the heterotic string theory  in arbitrary $d$ spacetime dimensions.  

In the previous section, we have solved the Boltzmann equation but have not analysed the cosmological equations of motion. In the context of string theory, however, flatness of the Universe at the quantum level should be taken into account with scrutiny. For this purpose, we will consider classical backgrounds in Minkowski spacetime, where supersymmetry is spontaneously broken at a scale $M$. The motivations  for this choice are the following: First, there is no (large) cosmological constant to start with at tree level, which  would have no chance to be cancelled by quantum effects in a perturbative regime. Second, the effective potential generated by loop corrections is of order $M^d$, which can be much lower than $M_s^d$, where $M_s\equiv 1/\sqrt{\alpha'}$ is the extremely large string scale and $\alpha'$ the string tension.  

In a string setup, and actually in supergravity, the theories where supersymmetry is spontaneously broken at the classical level in flat space are referred to as  ``no-scale models''~\cite{noscale}. The reason for this is that the scale $M$ is actually a scalar field, which turns out to be a flat direction of the tree-level potential. Hence, the vev of $M$  is undetermined at the classical level. As said before, this is however no more true at the quantum level. Switching on finite temperature, a thermal effective potential is generated, whose zero-temperature (Coleman--Weinberg) contribution is of order $M^d$, while the free energy part is of order $T^d$. Because the quantum thermal potential sources gravity in the Einstein equations, the  static background is not a solution anymore, but flat Friedmann-Lema\^itre-Robertson-Walker evolutions do exist. From this point of view, the cosmological  character of the Universe follows from pure quantum/thermal effects. 

At 1-loop, the thermal effective potential depends only on the mass spectrum at tree level. The latter depends on the vev of the so-called moduli fields, which are all scalar fields (including $M$) that are flat directions of the classical potential. They are the counterpart of the marginal deformations of the conformal field theory on the string worldsheet. The key point is that in heterotic string, massive states can become massless at special points in moduli space. In the following, we will consider the case where this arises when the radius $R_d$ of some compact (internal) direction takes the value $R_d=1$.\footnote{We express all dimensionful quantities in string units, with the convention $\alpha'=1$.}  To make contact with the previous sections, we introduce a canonical scalar field $\phi$ in terms of which we have $R_d\equiv e^\phi$, and write the mass that may vanish as  
\be
\left|R_d-{1\over R_d}\right|\simeq 2|\phi| \, .
\ee
Given these generalities, our aim is to specify some heterotic string background such that the \mbox{1-loop} thermal effective potential admits a temperature-dependent vacuum satisfying the following conditions:
\be
\begin{aligned}
&\mbox{High temperature : }\quad &\langle \phi\rangle=0 \, , \\
&\mbox{Low temperature : }\quad &\langle \phi\rangle\neq 0 \, .
\end{aligned}
\ee
In that case, some string states, interpreted as DM candidates, are initially participating to the thermal radiation. However, as the universe expands and cools, a phase transition  takes place and the condensation of $\phi$ induces a mass for the DM particles. Notice that in the present context, the scalar $\phi$ is a modulus, which means that $m_0(\phi)$ defined in Eq.~(\ref{m00}) vanishes identically. As a result, the thermal effective potential at 1-loop is perfectly consistent and it is not necessary to supplement it with higher-loop contributions. 

Our starting point is the heterotic string compactified on a $(10-d)$-dimensional torus, with 2 factorized circles of radii $R_d$ and $R_9$, 
\be
S_E^1(R_{0})\times \mathbb{R}^{d-1}\times  S^1(R_d) \times T^{8-d}\times  S^1(R_9)\, .
\ee
In order to implement finite temperature, time is Euclidean and also compactified on a circle of radius $R_0$. The Matsubara excitations have momenta along the temporal circle given by 
\be
{m_0+{F\over 2}\over R_0}\, , \quad m_0\in \mathbb{Z}\, ,
\ee
where $F=0$ for bosons and 1 for fermions, while the temperature is 
\be
T={1\over 2\pi R_0}\, .
\ee
We implement the spontaneous breaking of supersymmetry \textit{via} a stringy version~\cite{SSstring1,SSstring2,SSstring3} of the Scherk-Schwarz mechanism~\cite{SS1,SS2}. In field theory in $d+1$ dimensions, this amounts to imposing distinct boundary conditions to superpartner fields along the extra dimension. Hence, the degeneracy of their  Kaluza-Klein (KK) \textit{i.e.} Fourier modes living in $d$ dimensions is lifted. In the string context, introducing the Scherk-Schwarz breaking along the circle $S^1(R_9)$, a  particular set of KK modes is of special interest to us. Their squared masses are given by 
\be
\left(R_d-{1\over R_d}\right)^2+\bigg({m_9+{ F+B_{9d}\, n_d\over  2}\over R_9}\bigg)^2\, , \quad m_9\in \mathbb{Z}\, , \quad n_d=\pm 1\, ,
\label{KKm}
\ee
where $B_{9d}$ is one of the internal  components of the antisymmetric tensor. For given quantum numbers $m_9$, $n_d$, the degeneracy of these states is equal to 8.  In this expression, $n_d$ is the number of times a string mode winds around the direction $S^1(R_d)$. In this setup, the mass gap between bosonic and fermionic superpartners with $F=0$ or 1 determines the scale of supersymmetry breaking, which we defined as 
\be
M={1\over 2\pi R_9}\, .
\ee 
From inspection of Eq.~(\ref{KKm}), we see that massless states arise at $R_d=1$, when $B_{9d}$ is integer. Moreover, the parity of  $B_{9d}$ determines whether they are bosons of fermions:
\be
\begin{aligned}
&B_{9d}\in2\mathbb{Z} &&\Longrightarrow \quad \mbox{$N_{\rm extra}=2\times 8$ bosonic degrees of freedom are massless}\, ,  \\
&B_{9d}\in2\mathbb{Z}+1&&\Longrightarrow \quad \mbox{$N_{\rm extra}=2\times 8$ fermionic degrees of freedom are massless}\, .
\end{aligned}
\ee

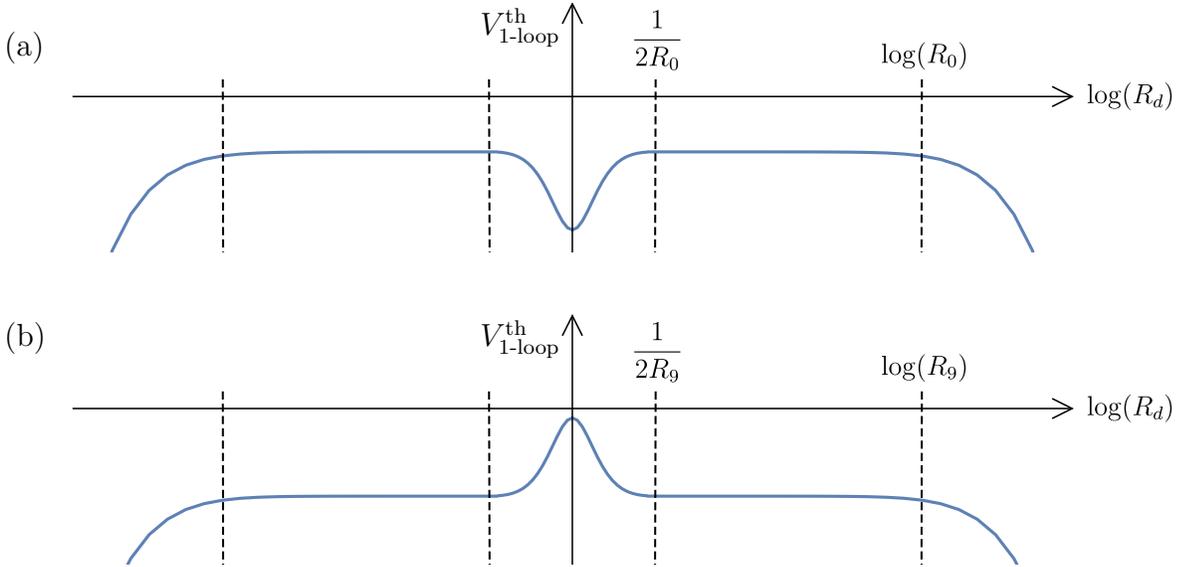
\begin{figure}
\vspace{.4cm}
\begin{center}
\includegraphics[trim=0cm 8.5cm 0cm 0cm,clip,scale=0.58]{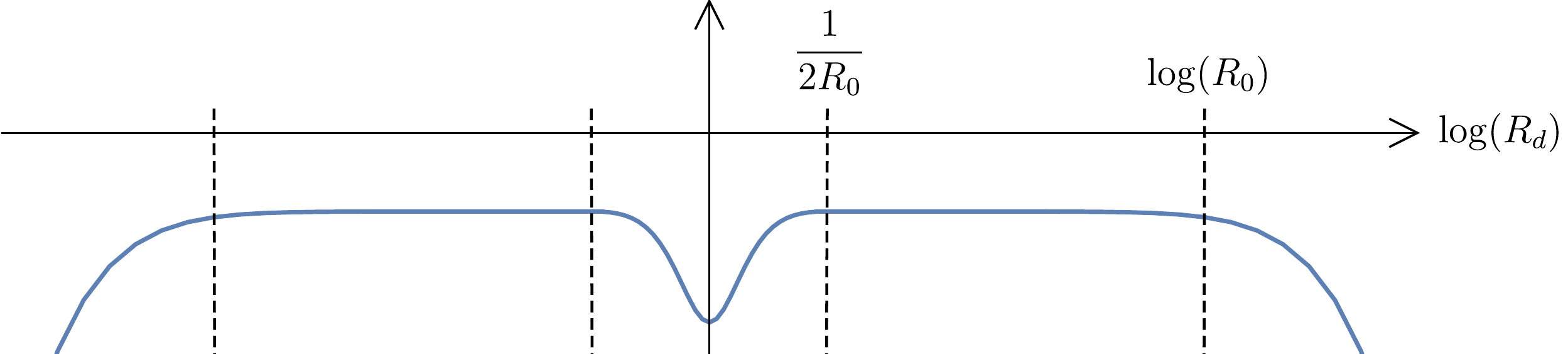}

\vspace{.8cm}
\includegraphics[trim=0cm 8.5cm 0cm 0cm,clip,scale=0.58]{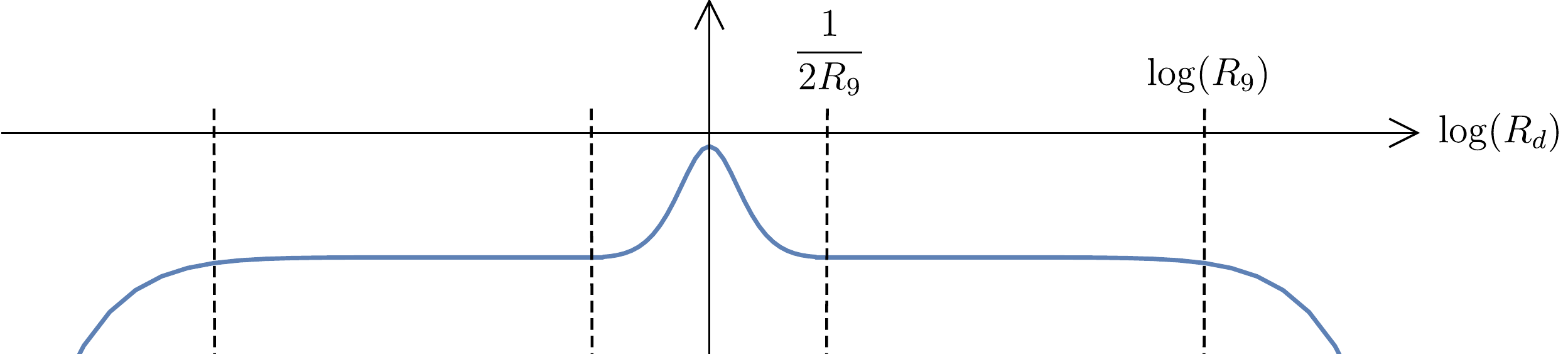}\end{center}
\begin{picture}(0,0)
\put(180,228){$\Vone^{\rm th}$}
\put(180,110){$\Vone^{\rm th}$}
\put(0,220){(a)}
\put(0,110){(b)}
\end{picture}
\vspace{-0.7cm}
\caption{\footnotesize \em Qualitative shapes of the 1-loop effective potential at finite temperature as a function of $\phi\equiv \log R_d$. The scalar can be {\rm (a)} massive or {\rm (b)} tachyonic.}
\label{pot}
\end{figure}

When the temperature and supersymmetry breaking scale are lower than then string scale, the expression of the 1-loop effective potential at finite temperature is dominated by the contributions of the light states that are the KK modes propagating along the internal compact directions $S^1(R_d)\times S^1(R_9)$. Expanding around $\phi=0$, one obtains~\cite{SFOstring}
\be
\begin{aligned}
\Vone^{\rm th}(T,\phi,z)=&\;T^4\Big[-(N_{\rm F}+N_{\rm B})f^{(d)}_T(z)+(N_{\rm F}-N_{\rm B})f^{(d)}_V(z)\Big] \\
&+{\phi^2\over \pi T^2}N_{\rm extra}\Big[f^{(d-2)}_T(z)+(-1)^{B_{9d}}f^{(d-2)}_V(z)\Big]+{\cal O}(\phi^4),
\end{aligned}
\ee
where $N_F$, $N_B$ are the total numbers of massless fermionic and bosonic degrees of freedom. In this expression, we have substituted the dependence on the supersymmetry breaking scale $M$ by that of a new variable  $z$ related to the ratio 
\be
 {M\over T}\equiv e^z\, , 
\ee
and we have defined the functions
\be
f^{(d)}_T(z)=\frac{\Gamma\!\left(\frac{d+1}{2}\right)}{\pi^{\frac{d+1}{2}}}\sum_{\tilde k_{0},\tilde k_{9}\in\Z}\frac{e^{dz}}{\left[e^{2z}(2\tilde k_{0}+1)^2+(2\tilde k_{9})^2\right]^{\frac{d+1}{2}}}\, , \quad f^{(d)}_V(z)\equiv e^{(d-1)z}f^{(d)}_T(-z)\, .
\ee
In the case where the extra massless states occurring at $\phi\equiv \log R_d=0$ are bosons, the scalar $\phi$ is massive whatever the temperature. The qualitative shape of the potential is shown in Fig~\ref{pot}a,   
where the vev $\langle \phi\rangle=0$ appears as a local minimum. On the contrary, when the $N_{\rm extra}$ massless modes are fermionic, the sign of the squared mass of $\phi$ depends on the ratio $M/T$. When $T>M$, the scalar is massive and the potential is as displayed in Fig~\ref{pot}a, while for $T<M$ it is tachyonic and globally the shape of the potential is as shown in Fig~\ref{pot}b. Therefore, for the destabilization of $\phi$ to be possible, we take $B_{9d}$ odd. 

However, for the phase transition to take place dynamically, the dependence of the thermal effective potential on the ratio $M/T$ does matter. It turns out that  when the massless spectrum satisfies the  condition
\be
0<{N_{F}-N_{B}\over N_{F}+N_{B}}<\frac{1}{2^{d}-1}\, ,
\ee
$\Vone^{\rm th}$ admits a minimum at some $\langle M/T\rangle$. When $\langle M/T\rangle>1$, the scenario expected to take place is the following: 
\begin{itemize}
\item Assuming initial conditions at some time $t_i$ such that $T(t_i)>M(t_i)$, the scalar $\phi$ is stabilized at $\langle \phi\rangle=0$, the $N_{\rm extra}$ fermionic degrees of freedom are massless,  and the ratio $M(t)/T(t)$ starts evolving. 
\item Because  $\langle M/T\rangle>1$,  $M(t)/T(t)$ eventually exceeds the value 1, implying the scalar $\phi$ to become tachyonic. 
\item $\phi$ is then destabilized, rolls along the bump of the potential and reaches one of the plateaus shown in Fig.\ref{pot}b, where it finally converges to a constant, due to the friction arising from the expansion of the universe. 
\item Throughout the evolution of $R_d(t)$, the $N_{\rm extra}$ fermionic states acquire mass and they freeze out at some instant $t_{\rm FO}$ defined by the condition  
\be
{\left| R_d(t_{\rm FO}) -\displaystyle {1\over R_d(t_{\rm FO})}\right|\over T(t_{\rm FO})}=\kappa\, .
\ee
\end{itemize}

To show that the above picture is valid, we have numerically simulated in explicit heterotic string models the cosmological evolutions of a minimal set of degrees of freedom in a homogeneous, isotropic and flat four-dimensional universe~\cite{SFOstring}. To be specific, we have taken into account the dynamics of the scale factor $a(t)$, the temperature $T(t)$, the scale of supersymmetry breaking $M(t)$, the scalar $\phi(t)$ and the string coupling $g_s(t)$, which is a field.  We found that $\phi(t)$ initially converges towards the minimum of its well, with  damped oscillations. However, this behaviour lasts until the critical time $t_c$ such that  $M(t_c)=T(t_c)$, after which the scalar field is destabilized, descending the bump of the potential and converging to some final vev $\langle \phi\rangle\neq 0$. As long as the freeze out of the $N_{\rm extra}$ fermionic modes does not take place, the evolution is attracted to a critical solution (for $d=4$ in the simulation) such that~\cite{RDS1,RDS2,RDS3,RDS4,RDS5}
\be
M(t)= T(t) \left\langle {M/ T}\right\rangle \propto {1\over a(t)}\propto g_s(t)^{2{d-1\over d-2}}\, , \qquad \phi(t)=\langle\phi\rangle\, .
\ee
The latter is said ``Radiation-Like'', because the total energy density and pressure arising from $(i)$ the thermal bath of the infinite towers of KK modes along the Scherk-Schwarz direction and $(ii)$ the coherent motion of $M(t)$ satisfy the state equation of pure radiation, 
\be
\rho_{\rm tot}=(d-1)P_{\rm tot}\, .
\ee
Of course, when DM freezes out, this evolution is modified when the component of the dust in the energy density of the universe starts to dominate.   

For completeness, we mention that in the string context we have reviewed, the issue of the stabilization of the string coupling $g_s(t)$ at later times must be  addressed. There are several reasons for that. First, when the gravitational kinetic term is in canonical (Einstein-Hilbert) form, all mass scales of the theory are dressed by a factor $g_s^{2d\over d-2}$. Hence, for the SM mass spectrum in realistic models to be time-independent, the field $g_s(t)$ must be stabilized. Second, if $g_s$ was not acquiring a mass, it would induce a long range force that would imply a violation of the experimental bounds on the validity of the Equivalence Principle.


\section*{Acknowledgements}

The  research  activities  of  L.H. are supported in part by the Department of Energy under Grant DE-FG02-13ER41976 (de-sc0009913). The work of H.P. is partially supported by the Royal-Society/CNRS International Cost Share Award IE160590.  This work was partially performed at the Aspen Center for Physics, which is supported by National Science Foundation grant PHY-1607611, as well as at the CERN Theory Department. This work was made possible by Institut Pascal at Universit\'e Paris-Saclay with the support of the P2I and SPU research departments and the P2IO Laboratory of Excellence (program ``Investissements d'avenir'' ANR-11-IDEX-0003-01 Paris-Saclay and ANR-10-LABX-0038), as well as the IPhT.



\end{document}